\documentclass[aps,prl,twocolumn]{revtex4-2}

\usepackage{amsmath,amssymb,amsthm}
\usepackage{mathtools}
\usepackage{physics}
\usepackage{hyperref}
\usepackage{microtype}
\usepackage{booktabs}
\usepackage{xcolor}
\usepackage{placeins}
\usepackage{subcaption}

\hypersetup{colorlinks=true, linkcolor=blue, citecolor=blue, urlcolor=blue}

\newcommand{\cO}{\mathcal{O}}
\newcommand{\cF}{\mathcal{F}}

\newcommand{\ep}{\varepsilon}
\newcommand{\AK}[1]{{\color{red} \bf }}
\newcommand{\AT}[1]{{\color{ cyan} \bf }}

\begin{document}

\title{Truncated Polyakov bootstrap for BCFTs:\\ Neumann-Dirichlet flow and Ising special transition}

\author{Kaushik Kangsabanik 
}
\author{Apratim Kaviraj}
\author{Astha Tiwari}

\affiliation{Department of Physics, Indian Institute of Technology - Kanpur, Kanpur 208016, India}


\begin{abstract}
\noindent

We extend the recently proposed truncated Polyakov bootstrap to boundary conformal field theories (BCFTs). The formalism uses the analytic functional framework to find approximate solutions of BCFT crossing. We show that a generic nonperturbative solution can be identified by relating it to a family of deformations around generalized free fields (GFF) with a fixed boundary condition. We show this by tracking the flow of BCFT data from GFF with Neumann boundary condition to that with Dirichlet. The same method is then used for the 3d Ising special transition, using the perturbative Wilson-Fisher description as the deformation to identify it. The BCFT data obtained are in remarkable agreement with existing lattice results where available, and are otherwise new.
\end{abstract}
\maketitle
\section{Introduction}
\label{sec:intro}

\vspace{-0.1cm}

Near a continuous phase transition physical observables demonstrate critical behavior which in the infinite volume limit are described by a conformal field theory. A real system, however, has a finite volume that  introduces a large length scale that significantly affects the critical properties.  These finite size effects are determined by the critical phenomena near a surface or boundary of the system, the physics of which is nicely captured by a boundary conformal field theory (BCFT). 

A BCFT is an example of a codimension~1 conformal
defect. It is a $d$-dimensional CFT in presence of a boundary such that the theory on boundary is also conformal. Away from the boundary, the conformal symmetry of the bulk is approximately
preserved, but close to it the correlation functions get corrected according
to how bulk fields are coupled to the boundary. So the conformal data of the
boundary CFT depends on both the OPE data of the bulk and the boundary
conditions.

Given a bulk CFT and its boundary conditions the BCFT data can be obtained analytically in perturbation theory. A lot of analytic results and related tools have been developed for boundaries on the Wilson-Fisher, $O(N)$ and other interesting bulk theories \cite{Diehl1981FieldtheoreticalAT, REEVE1981237, PhysRevLett.45.1581, PhysRevB.11.4533}, see \cite{Bissi2018, Dey:2020lwp, Giombi:2019enr, Hu:2025yrs, Guo:2025edk, Shpot:2019iwk, Gimenez-Grau:2020jvf, Guo:2026vmq, Diatlyk:2026eta, sun2026analyticbootstraponboundary, guo2026boundaryanomalousdimensionsbcft} for recent works. On the nonperturbative front, a lot of important results have been obtained from  lattice-based Monte-Carlo analysis \cite{Przetakiewicz2025}. More recently the Fuzzy sphere approach has also been very successful \cite{Zhou_2025, Dedushenko:2024nwi, Feng:2026iii}.


While the above techniques are model-dependent, the BCFT data are universal and can be fixed from axiomatic principles just like ordinary CFTs.  
The BCFT bootstrap formalism \cite{Liendo_2012} takes this approach. 
The BCFT crossing equation relates bulk and boundary conformal data. But unlike ordinary 4-point crossing it faces a major challenge: in 
general it cannot be framed as a positive semi-definite
problem, even if theory is unitary. So the space of solutions
cannot be isolated by bounds. The \emph{truncated bootstrap} is therefore the
more natural tool for BCFTs. The idea there is to fix a part of the spectrum
in order to find the rest of it.

In \cite{Kangshabanik:2026onk} a new formalism --- the
\emph{truncated Polyakov bootstrap}, was proposed, and demonstrated for 1d CFTs. It is a truncation scheme based on  the Polyakov bootstrap (PB) or analytic functional framework for ordinary CFT 4-point crossing. The main idea is to connect a
general crossing solution to a generalized free theory (GFF) by starting
with a small deformation and increasing it in small steps. A
family of solutions is obtained this way by requiring the fewest possible
states in the OPE. 
For unitary deformations such families can be linked to extremal (i.e. bound saturating) theories \cite{Zan:2019fkw, Ghosh:2025sic}, but with truncated PB the solutions need not be unitary i.e. with positive semidefinite OPE coefficients.  This gives a way to identify any solution based on what deformation connects it
smoothly to GFF.

In this paper we extend the truncated PB to the BCFT
framework. The required tools are the analytic functional sum rules,
developed in \cite{Kaviraj2019, Mazac:2018biw}. Our first application is to identify a family of solutions connecting a bulk GFF with
 Neumann boundary conditions to that with Dirichlet boundary
conditions. This is the closest analogue of an extremal solution family in BCFT crossing (without positivity). 
In the AdS/BCFT holographic context this can be thought of as a flow  triggered by a deformation on an end-of-the-world brane. See \cite{Behan_2020, Behan:2021tcn, Banerjee:2026gce} for related works.

\begin{figure}[t]
    \centering
    \includegraphics[width=\columnwidth]{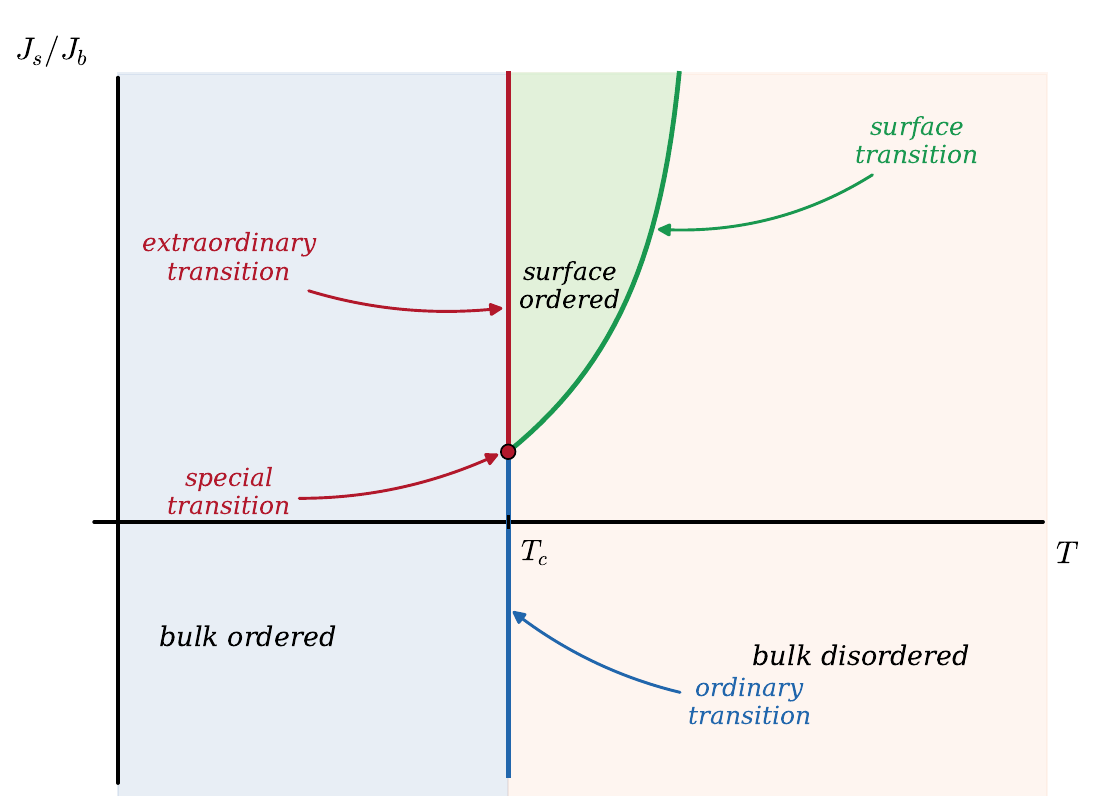}
    \caption{Phase diagram of 3d Ising with a boundary}
    \label{fig:bcft-Ising}
\end{figure}

The second, and more important, goal of this paper is to find the 3d Ising
\emph{special transition} BCFT. When the 3d Ising spin system is at the critical point ($T=T_c$) its boundary is associated to three different transitions: ordinary, special and extraordinary, see Fig. \ref{fig:bcft-Ising}. The ordinary and extraordinary transitions \cite{Padayasi:2021sik, Dedushenko:2024nwi} are not explored in this paper. The special transition is an unstable point  reached by fine-tuning the ratio $J_s/J_b$ where $J_s$ and $J_b$ are the surface and bulk couplings between Ising spins respectively \cite{Cardy1984, Cardy_1996}. Perturbatively it is described by the Wilson-Fisher CFT with Neumann boundary conditions.
We use the leading boundary scaling dimension input from  Monte-Carlo \cite{Przetakiewicz2025} and
find a set of BCFT data, most of which are unknown in the literature.


\section{BCFT bootstrap review}
In CFTs, the one-point function is always trivial because of translation and dilatation symmetry. On adding $p$-dimensional defects, the conformal symmetry group breaks into SO$(p+1,1)\times $SO$(q)$, where $q=d-p$ is the co-dimension of the defect. Then the form of one point function gets fixed in terms of the radial distance to the defect. For a boundary (co-dimension 1) the conformal group breaks to the subgroup SO$(d,1)$, whereas the operators living on the boundary transform under the full symmetry of $d-1$ dimensional CFTs.  

Consider a CFT on a Euclidean half-space, with a generic (bulk) point labelled as $x = (\vec{x}, r)$, where $\vec{x}$ denotes the $(d-1)$ components along the boundary and $r := x^d$ is the distance from the boundary. So the boundary itself is  at $r=0$.
If we consider a scalar bulk operator $\cO$ of dimension $\Delta$ then its one-point function takes the form \begin{equation}\label{1ptfn}
  \langle \cO(x) \rangle = \frac{a_\cO}{(2\,r)^{\Delta}}\,,
\end{equation}
where $a_\cO$ is the one-point coefficient.
In boundary CFTs, only scalar operators can have non-zero one-point function.
We use $\Delta$ for bulk operator dimensions and $\hat{\Delta}$ for boundary operator dimensions throughout.

Now take two bulk scalars $\phi(x_1)$ and $\phi(x_2)$. Their two-point function depends on a single cross-ratio $z$ which, in the convention of \cite{Kaviraj2019}, is given by
\begin{equation}
  z = \frac{4\,r_1\,r_2}{|x_1-x_2|^2 + 4\,r_1 r_2}\,,
  \label{eq:zdef}
\end{equation}
and related to the more commonly used cross-ratio $\xi = |x_1-x_2|^2/(4 r_1 r_2)$ by $\xi = (1-z)/z$. In terms of $z$ the two-point function takes the form
\begin{equation}\label{phiphi}
  \langle \phi(x_1)\,\phi(x_2) \rangle
  = z^{\Delta_\phi}\, \frac{\cF(z)}{(r_1\,r_2)^{\Delta_\phi}}\, .
\end{equation}
Here, and in the rest of the paper, we take both external bulk operators to have the same scaling dimension $\Delta_\phi$.

\paragraph{Bulk expansion:} The two-point function admits a decomposition in terms of the operators appearing in the OPE of $\phi\times\phi$ of the bulk theory. This is the same OPE as the CFT$_d$ without the boundary, and written as
\begin{equation}
\phi(x)\phi(0)
=
\sum_{\Delta,\ell} \lambda_{\Delta,\ell}\,C(x,\partial)\,\mathcal{O}_{\Delta,\ell}(0)
\, .
\end{equation}
Here $\Delta$ and $\ell$ denote the dimension and spin respectively for each primary operator $\mathcal{O}_{\Delta,\ell}$, and $\lambda_{\Delta,\ell}$ denotes its OPE coefficient. The $C(x,\partial)$ accounts for the descendents and is fixed by conformal invariance. 
Now OPEs are local properties of a CFT and do not depend on the geometry of the manifold the theory is on. So the above form remains unchanged in the presence of a boundary. We may use it for the 2-point function in \eqref{phiphi}. This gives an expansion
\begin{equation}
z^{\Delta_\phi}\,\cF(z) = \sum_{\Delta} a_{\Delta}\, \lambda_{\Delta}\, G_{\Delta}(z|\Delta_\phi),
\end{equation}
where $a_\Delta$ is the bulk one-point coefficient from \eqref{1ptfn} and $\lambda_\Delta:=\lambda_{\Delta,0}$. Notice that only scalar operators are present in this expansion.

The bulk expansion is natural when the boundary is far away from $x_1, x_2$, in other words $z\approx 1$. Following the conventions of \cite{Kaviraj2019}, the bulk conformal block is $G_{\Delta}(z|\Delta_\phi) = (1-z)^{\frac{\Delta}{2}-\Delta_\phi}\, {}_2F_1\!\left(\frac{\Delta}{2},\, \frac{\Delta}{2}-\ep,\, \Delta-\ep;\, 1-z\right)$
with $\ep \equiv \frac{d-2}{2}$ \cite{McAvity_1995}. From here on we will define  $(a\lambda)_\Delta := a_\Delta \lambda_\Delta$ and simply call it the OPE coefficient.

\paragraph{Boundary expansion:}
Close to the
boundary, the bulk operator can be expressed as a sum over boundary primary operators $\hat{\mathcal{O}}$ and their descendants 

\begin{equation}
\phi(x)
=
\sum_{\hat{\mathcal{O}}}
\mu_{\hat{\Delta}}\,
\hat{C}_{\hat{\Delta}}(r,\partial_{\vec{x}})\,
\hat{\mathcal{O}}(\vec{x})
\, .
\end{equation}
Here the $\hat{C}_{\hat{\Delta}}$ function captures the descendent contributions and is fixed by conformal symmetry. 
Writing the two point function using the above expansion gives \cite{McAvity_1995}

\begin{equation}
z^{\Delta_\phi}\,\cF(z) = \sum_{\hat{\Delta}} \mu_{\hat{\Delta}}^2\, \widehat{G}_{\hat{\Delta}}(z|\Delta_\phi).
\end{equation}
Here we introduced the boundary conformal block  given by $\widehat{G}_{\hat{\Delta}}(z|\Delta_\phi) = z^{\hat{\Delta}-\Delta_\phi}\, {}_2F_1\!(\hat{\Delta},\, \hat{\Delta}-\ep,\, 2\hat{\Delta}-2\ep;\, z)$\,.  We will refer to $\mu_{\hat{\Delta}}^2$ as  boundary operator expansion (BOE) coefficient.


\paragraph{BCFT crossing:} Since both decompositions describe the same two-point function, equating them gives the boundary bootstrap crossing equation
\begin{equation}
\sum_{\hat{\Delta}} \mu_{\hat{\Delta}}^2\, \widehat{G}_{\hat{\Delta}}(z|\Delta_\phi) = \sum_{\Delta} (a\lambda)_{\Delta}\, G_{\Delta}(z|\Delta_\phi)\,.
\label{eq:crossing}
\end{equation}
Here the bulk dimensions $\Delta$ are treated as known and the rest, i.e. $(a\lambda)_\Delta, \mu^2_{\hat\Delta}, \hat\Delta$,  as unknowns.

\paragraph{Boundedness at large $z$:} A physical BCFT 2-point function can be shown to be bounded at large $z$ \cite{Mazac:2018biw, Kaviraj2019}
\begin{equation}\label{Regge}
    z^{-\eta}\mathcal{F}(z)\stackrel{z\to \infty}{\to} 0\,, \ (\eta>0)\,.
\end{equation}
This property is analogous to the boundedness of an ordinary CFT 4-point function at large cross-ratios (Regge limit).

\section{Generalized free fields}
Generalized free field (GFF) theories provide a simple yet important example of BCFTs. 
A bosonic GFF BCFT corresponds to the holographic dual of the theory of a massive free scalar $\Phi$ in AdS$_{d+1}$, in presence of a $d$-dimensional brane $\text{AdS}_d$ at $r=0$.
The 2-point function of $\phi$, the dual of $\Phi$, is obtained by Wick contractions: 
\begin{equation}
\begin{split}
\langle \phi(x_1) \phi(x_2) \rangle 
&= \frac{1}{\left(|\vec{x}_{12}|^2 + (r_1 - r_2)^2\right)^{\Delta_\phi}} \\
&+ \frac{\nu}{\left(|\vec{x}_{12}|^2 + (r_1 + r_2)^2\right)^{\Delta_\phi}}\,,
\end{split}
\end{equation}
where  $|\vec{x}_{12}|^2 + (r_1-r_2)^2 = |x_1-x_2|^2$. Separating the prefactor gives
\begin{equation}
\cF(z) = \frac{1}{(1-z)^{\Delta_\phi}} + \nu.
\end{equation}
where $\nu=+1$ and $\nu=-1$ for Neumann, Dirichlet boundary conditions, respectively for the AdS field $\Phi$ on the brane. In the BCFT the bulk channel scaling dimensions are  $\Delta_n=2 \Delta_\phi+2n$, i.e. the double twist dimensions from the usual OPE of a GFF field $\phi$ without boundary. In the boundary channel the scaling dimensions are given by $\hat\Delta_n^\nu=\Delta_\phi+2n+(\nu+1)/2$  specifically $\Delta_\phi+2n$ for the Neumann Boundary conditions (NBC), and $\Delta_\phi+2n+1$  for Dirichlet (DBC). All the OPE and BOE  coefficients are known exactly : 
\begin{align}
(a\lambda)_{\Delta^\nu_n}
&=
\nu (-1)^n
\frac{
(\Delta_\phi)_n
\left(2\Delta_\phi-1-\epsilon+2n\right)_{-n}
}{
\left(\Delta_\phi-\epsilon+n+1\right)_{-n}\, n!
},
\label{gff1}\\
\mu^2_{\hat\Delta_n^\nu}&=
\frac{
(\Delta_\phi)_{2n+\frac{1-\nu}{2}}
\left(\Delta_\phi-\epsilon-\frac{\nu}{2}+2n\right)_{-n}
}{
2^{2n-\frac{1+\nu}{2}}
\left(2n+\frac{1-\nu}{2}\right)!
\left(\Delta_\phi-\epsilon+n\right)_{-n}
}.\label{gff2}
\end{align}
Below we use the notation $(a\lambda)^{\text{gff}}_{\nu,n}:=(a\lambda)_{\Delta^{\nu}_n}$ and $\mu^{\text{gff}}_{\nu,n}:=\mu^2_{\hat\Delta_n^\nu}$ to refer to them.

\section{BCFT analytic functionals }
\label{sec:bcftfunctionals}
Let us review the analytic functional approach to BCFT bootstrap briefly. In ref.~\cite{Kaviraj2019} it was shown that both the bulk and boundary blocks admit a decomposition as follows:
\begin{align}
  G_{\Delta}(z|\Delta_\phi) &= \sum_{n=0}^{\infty} \theta_n(\Delta)\, G_{\Delta_n}(z|\Delta_\phi)
  + \sum_{n=0}^{\infty} \alpha_n(\Delta)\, \widehat{G}_{\hat{\Delta}_n}(z|\Delta_\phi) \notag \\
  &\qquad + \sum_{n=1}^{\infty} \beta_n(\Delta)\, \partial_{\hat{\Delta}} \widehat{G}_{\hat{\Delta}_n}(z|\Delta_\phi)\,,
  \label{eq:bulkblockbasis} \\
  \widehat{G}_{\hat{\Delta}}(z|\Delta_\phi) &= \sum_{n=0}^{\infty} \theta_n(\hat{\Delta})\, G_{\Delta_n}(z|\Delta_\phi)
  + \sum_{n=0}^{\infty} \alpha_n(\hat{\Delta})\, \widehat{G}_{\hat{\Delta}_n}(z|\Delta_\phi) \notag \\
  &\qquad + \sum_{n=1}^{\infty} \beta_n(\hat{\Delta})\, \partial_{\hat{\Delta}} \widehat{G}_{\hat{\Delta}_n}(z|\Delta_\phi)\,.
  \label{eq:bryblockbasis}
\end{align}
In both the equations, the blocks appearing on the r.h.s.\ are basis vectors. Their coefficients in this basis follow from the dual linear functionals $\theta_n$, $\alpha_n$, and $\beta_n$. E.g.\ $\theta_n(\Delta):=\theta_n[G_{\Delta}(z|\Delta_\phi)]$ and $\theta_n(\hat{\Delta}):=\theta_n[\widehat{G}_{\hat{\Delta}}(z|\Delta_\phi)]$ are the actions of the functional $\theta_n$. The $\alpha_n(\Delta)$ and $\beta_n(\Delta)$  are defined analogously.

Let us demonstrate the action of the functionals with the example of $\beta_1$, which is given by
\begin{align*}
\beta_1[F(z)] &= \int_1^{\infty} dz\; h(z|\Delta_\phi)\,\mathrm{Im}[F(z)]\,, \\
h(z|\Delta_\phi) &= \frac{\Gamma(\Delta_\phi+2)}{\pi^{3/2}}\, z^{-\frac{5}{2}}\,
{}_2\tilde{F}_1\!\left(\mbox{$-\frac{1}{2},\frac{3}{2};\Delta_\phi+\frac{3}{2};\frac{1}{z}$}\right)\,.
\end{align*}
We review the explicit form of the other functionals in the appendix. Their construction depends crucially on the analytic properties of the blocks and also the correlator. The existence of the functional basis shown in Eq.~\eqref{eq:bulkblockbasis}-\eqref{eq:bryblockbasis} is particular to correlators with the boundedness property \eqref{Regge}.

The actions of the analytic functionals on the BCFT bootstrap equation~\eqref{eq:crossing} give the functional sum rules 
\begin{align}
 &  A_n: \quad \sum_{\hat{\Delta}} \mu^2_{\hat{\Delta}}\,\alpha_n(\hat{\Delta})
  - \sum_{\Delta}(a\lambda)_\Delta\,\alpha_n(\Delta)=0, \label{eq:fbe_a} \\
  &  B_n: \quad  \sum_{\hat{\Delta}} \mu^2_{\hat{\Delta}}\,\beta_n(\hat{\Delta})
  - \sum_{\Delta}(a\lambda)_\Delta\,\beta_n(\Delta)=0, \label{eq:fbe_b} \\
  &  T_n: \quad  \sum_{\hat{\Delta}} \mu^2_{\hat{\Delta}}\,\theta_n(\hat{\Delta})
  - \sum_{\Delta}(a\lambda)_\Delta\,\theta_n(\Delta)=0. \label{eq:thetasumrule}
\end{align}
The sum rules can also be seen by using \eqref{eq:bulkblockbasis} and \eqref{eq:bryblockbasis} on \eqref{eq:crossing} and swapping $n$ with $\Delta$ and $\hat\Delta$ sums. 

The functional basis has orthogonality, meaning each functional, say $\theta_n$, acting on Eq.~\eqref{eq:bulkblockbasis} (or \eqref{eq:bryblockbasis}) picks out only the corresponding coefficient $\theta_n(\Delta)$ (or $\theta_n(\hat\Delta)$) and eliminates the other terms.  Orthogonality can be compactly expressed as follows: 
\begin{align}\label{ortho1}
    &\omega_{n}(\Delta_m)=\delta_{\omega\theta}\delta_{mn}\,,\ \omega_{n}(\hat{\Delta}_m)=\delta_{\omega\alpha}\delta_{mn}\,, \nonumber\\
    &\partial_{\Delta}\omega_{n}(\hat{\Delta}_m)=\delta_{\omega\beta}\delta_{mn}+c_{\omega,n}(\Delta_\phi)\delta_{m0}\,.
\end{align}
Above $\omega=\theta, \alpha, \beta$. The expressions for $c_{\omega,n}$ are known in closed form. In addition to these properties, the action of the functionals on the bulk identity: $\alpha_n(0)=\mu_{+,n}^{\text{gff}}$, $\theta_n(0)=(a\lambda)_{+,n}^{\text{gff}}$. This amounts to an interesting feature: the OPE data of the GFF spectrum \eqref{gff1}-\eqref{gff2} with Neumann B.C. ($\nu=1$) is a trivial solution of the sum rules~\eqref{eq:fbe_a}-\eqref{eq:thetasumrule}. Therefore $\alpha_n, \beta_n, \theta_n$ are called the \textit{Neumann boundary functionals}.

The sum rules also manifest small deformations around the $\nu=1$ GFF solution. E.g.: if we fix the bulk dimension at GFF but deform the BCFT data
\begin{align}
    \hat{\Delta}_n &= \Delta_\phi+2n+\gamma_n\,, \notag  \\
    \mu^2_{\hat{\Delta}_n} &= \mu^{\text{gff}}_{+,n}+\delta_n\,, \notag\\
    (a\lambda)_{\Delta_n} &= (a\lambda)^{\text{gff}}_{+,n}+\rho_n\,, \label{eq:gffdeform}
\end{align}
where $\gamma_n,\delta_n,\rho_n \ll 1$, then these parameters are analytically solved by the sum rules as follows:
\begin{equation}
    \rho_n=-c_{\theta,n}\gamma_0\,,\ \delta_n=-c_{\alpha,n}\gamma_0\,,\ \gamma_n = -c_{\beta,n}\gamma_0\,.
    \label{eq:gammapert}
\end{equation}
This particular deformation can be thought of as a perturbation  by $g\,{\Phi}^2$ term on $\text{AdS}_d$ i.e. an interaction on the brane. The $\gamma_0$ is simply a replacement of $g$.

There is also a set of functionals that trivializes the GFF with Dirichlet B.C., see \cite{Kaviraj2019}. We will not use the Dirichlet functionals in this paper.

\section{Truncated Polyakov bootstrap}
\label{sec:results1}
The implementation of the truncated PB begins by introducing a loss function
\begin{equation}
    \mathcal{F} = \sum_{n=0}^{n_{\text{max}}} a_n A_n^2 + \sum_{n=1}^{n_{\text{max}}} b_n B_n^2 + \sum_{m=0}^{m_{\text{max}}} t_m T_m^2\,.
    \label{eq:lossfunction}
\end{equation}
Here $A_n, B_n, T_m$ denote the l.h.s. of the sum rules in \eqref{eq:fbe_a}-\eqref{eq:thetasumrule} and $a_n, b_n, t_m$ are their strengths which we discuss below. 
We will look for a global minima of the loss  in the multidimensional space of BCFT data. This will simultaneously minimize all the sum rules. Consider the BCFT data shown in \eqref{eq:gffdeform} - if all $\gamma_n, \delta_n, \rho_n$ are 0 then the sum rules are trivially satisfied and $\mathcal{F}$ is at a global minima. Now let us introduce a deformation, e.g. by setting $\gamma_0$ to a small value or adding a set of new operators in bulk or boundary channel with small coefficients - the minima of $\mathcal{F}$ will move to a new point with nonzero $\gamma_n, \delta_n, \rho_n$. We will locate this new minima by gradient descent. We assume that there is no other local minima  nearby \footnote{If there is an unphysical local minima it can be eliminated by adding more sum rules in $\mathcal{F}$ - they will not converge to 0. }.

We will increase the deformation from small to a finite value in small steps. At every step we will  minimize $\mathcal{F}$ starting from  the minima point of the previous step. This initialization or warm-starting is crucial as it shrinks the search space dramatically - giving us a unique minima at every step. 

At every step, if any of the sum rule starts off with a very small value it will not be accurately accounted for in the minimization process. This is why we include the strengths in the definition \eqref{eq:lossfunction}.  E.g.: if $\gamma_0=0.1$ and we initialize with the other corrections  zero, then $a_n$, $b_n$ or $t_m$ may be set close to $A_n^{-2}$, $B_n^{-2}$ and $T_m^{-2}$ respectively, evaluated at those values. The strengths can still be varied by keeping their order of magnitude fixed - which we use to estimate the error in results. 





\section{Neumann to Dirichlet flow}
The first numerical analysis in our approach is to extend the result~\eqref{eq:gammapert} by pushing $\gamma_0$ to larger values. The objective is to find a set $\{\gamma_n,\delta_n,\rho_n\}$ for a $\gamma_0>0$, by minimizing $\mathcal{F}$ by gradient descent. We will truncate the OPE data to have $n_{\text{max}}$ boundary operators and $m_{\text{max}}$ bulk operators. This ensures a controlled minimization both in the perturbative and non-perturbative regimes.

\begin{figure}[t]
  \includegraphics[width=0.9\columnwidth]{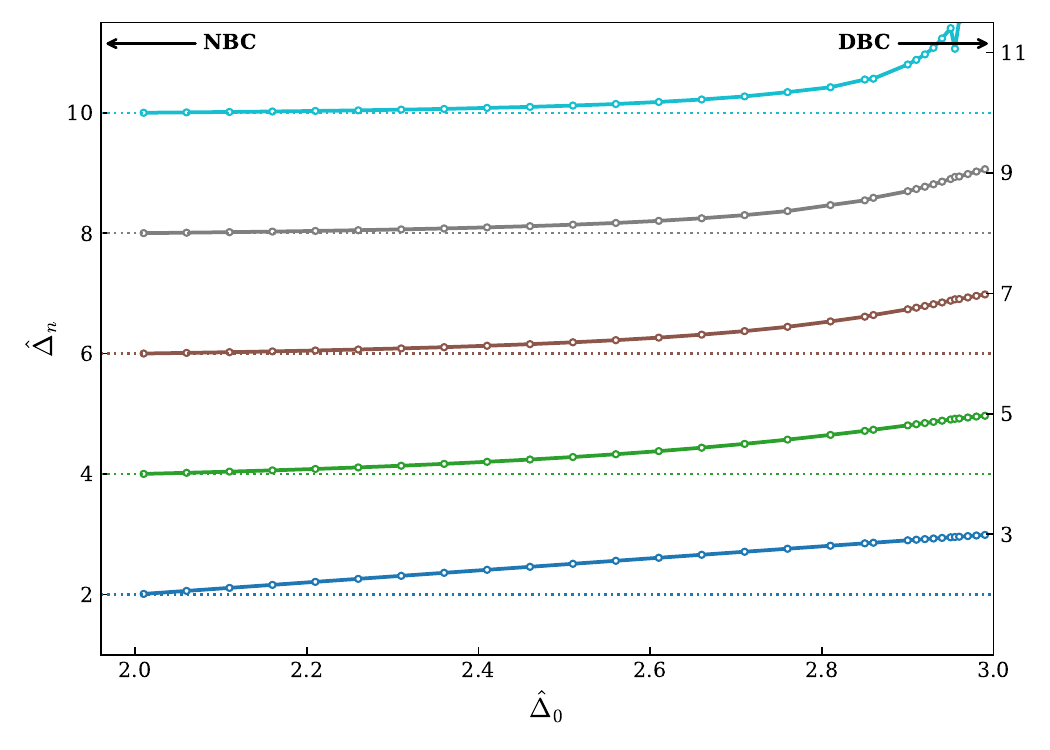}
  \caption{Solid lines: Flow of boundary operator dimensions $\hat{\Delta}_n$ 
  ($n = 0,1,2,3,4$) as a function of $\hat{\Delta}_0$.
  Dotted lines: 
  GFF with Neumann 
  boundary condition (NBC). On the right spectrum approaches that of Dirichlet boundary condition (DBC), 
  $\hat{\Delta}_n = \Delta + 2n + 1$.}
  \label{fig:deltaflow}
\end{figure}

\begin{figure}[t]
    \centering
    \includegraphics[width=\columnwidth]{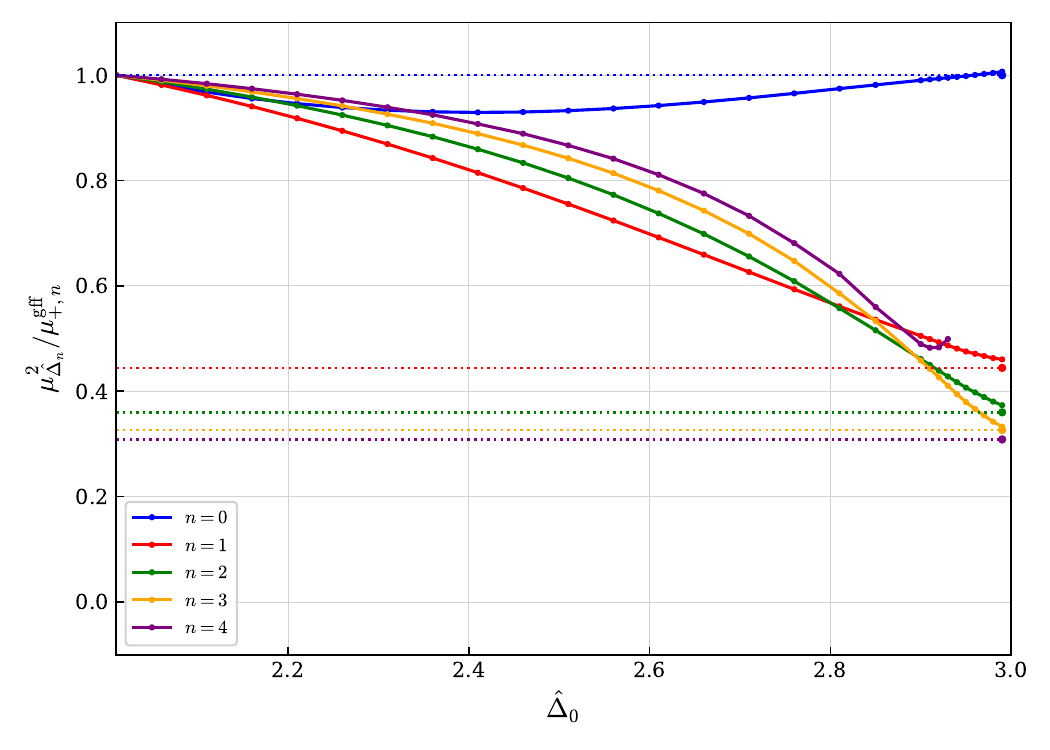}
    \caption{Solid lines: flow of $\mu^2_{\hat{\Delta}_n}/\mu^{\text{gff}}_{+,n}$ vs. $\hat{\Delta}_0$ (the $n=4$ line becomes unstable after $\hat{\Delta}_0\approx 2.91$). Dotted lines: $\mu^{\text{gff}}_{-,n}/\mu^{\text{gff}}_{+,n}$  i.e. the values at the DBC point.}
    \label{fig:BOE_ratio}
\end{figure}

\begin{figure}[t]
  \includegraphics[width=\columnwidth]{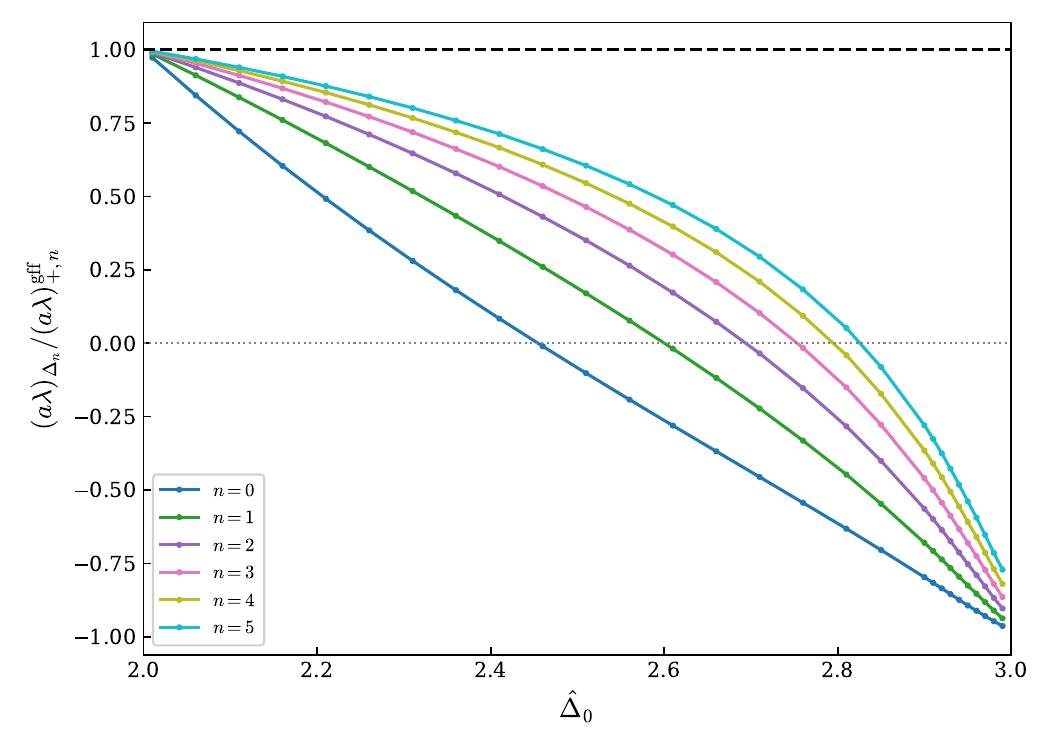}
    \caption{Flow of $(a\lambda)_{\Delta_n}/(a\lambda)_{+,n}^{\text{gff}}$ vs. $\hat{\Delta}_0$. For DBC $(a\lambda)_{-,n}^{\text{gff}}/(a\lambda)_{+,n}^{\text{gff}}=-1$ from \eqref{gff1}.}
  \label{fig:operatio}
\end{figure}

We have varied $\hat\Delta_0$ from $\Delta_\phi$ to $\Delta_\phi+1$, for $\Delta_\phi=2$. The bulk theory is always GFF, i.e.\ $\hat\Delta_n=2\Delta_\phi+2n$. The minimal spectrum (i.e.\ the BCFT data that minimizes $\mathcal{F}$) is shown in Fig. ~\ref{fig:deltaflow}, ~\ref{fig:BOE_ratio} and ~\ref{fig:operatio}. In Fig.~\ref{fig:deltaflow} we show the variation of $\hat{\Delta}_n$. The leftmost part in this plot matches approximately to the analytically obtained solution~\eqref{eq:gammapert}. On the right we see that $\hat{\Delta}_n$ are approaching $\Delta_\phi+2n+1$. Therefore, quite remarkably, the flow initiated by a ${\Phi}^2$ deformation on the AdS brane seems to reach the Dirichlet spectrum. The corresponding BOE coefficient $\mu_{\Delta_n}^2$ flow, having the same feature is shown in Fig. \ref{fig:BOE_ratio}. 




Figure~\ref{fig:operatio} shows the ratio $(a\lambda)_{\Delta_n}/(a\lambda)_{+,n}^{\text{gff}}$ of the 
bulk OPE coefficients. All of them start at 
unity at the NBC point and flow to $-1$ at the DBC point. From \eqref{gff1} one finds $(a\lambda)_{-,n}^{\text{gff}}/(a\lambda)_{+,n}^{\text{gff}}=-1$. Our result is nicely  consistent 
with this sign flip under the 
boundary condition change.

\section{3D Ising special transition}
We will now use the truncated PB to bootstrap the 3d Ising special boundary transition. For this we will fix the bulk CFT data and also the leading dimension on the boundary. However this may not be enough to locate the solution we want, as there may still be multiple BCFT data sets consistent with these inputs. To resolve this ambiguity we start with a perturbative description of the desired BCFT and find a family of solutions that smoothly connects to the target.

The special transition can be described with a Wilson-Fisher BCFT, i.e. the IR fixed point of the $\phi^4$ scalar theory in $d=4-\epsilon$ bulk dimension, in presence of a boundary with Neumann boundary conditions. The system has a global $Z_2$ symmetry. 

The lightest bulk field $\phi$ is a $Z_2$ odd operator.  In the perturbative setup (small $\epsilon$), its IR scaling dimension is $\Delta_\phi=1-\frac{\epsilon}{2}+\frac{\epsilon^2}{108}+O(\epsilon^3)$.
Consider its 2-point function $\langle\phi(x)\phi(0)\rangle$. It has the following bulk OPE \cite{Bissi_2019, sun2026analyticbootstraponboundary}:
\begin{align}\label{bulkWF}
    1  : &  \ \Delta=0\,, \hspace{1cm} (a\lambda)_{\Delta=0}=1\,,\nonumber\\
    \phi^2  : &  \   \Delta=\Delta_{\phi^2}=\mbox{$2-\frac{2\epsilon}{3}+\frac{19\epsilon^2}{162}+O(\epsilon^3)$}\,, \nonumber\\
    & \ (a\lambda)_{\Delta_{\phi^2}}=\mbox{$1+\frac{\epsilon}{6}-\frac{\epsilon^2}{162}+O(\epsilon^3)$}  \,.
\end{align}
Other operators are suppressed by higher orders of $\epsilon$ in the coefficient $(a\lambda)_\Delta$. Note that operators in the bulk channel have to be $Z_2$ even. In the boundary channel one has \cite{Bissi_2019, sun2026analyticbootstraponboundary}:
\begin{align}\label{bdyWF}
   \hat{\phi}: \ \  &\Delta=\hat{\Delta}_\phi=1-\mbox{$\frac{2}{3}\epsilon+\frac{5\epsilon^2}{324}+O(\epsilon^3)$}\,, 
 &\mu^2_{\hat\Delta_\phi}=2+O(\epsilon^3)\,.
\end{align}
Once again other operators are suppressed by higher powers of $\epsilon$. Note that in the boundary channel we can have only $Z_2$ odd operators. 

Notice that for $\epsilon=0$ i.e. in exact 4d the spectrum in \eqref{bulkWF} and \eqref{bdyWF} is that of free theory. In fact this is GFF with Neumann B.C.  with $\Delta_\phi=1$, where most of the OPE data go to zero from \eqref{gff1}-\eqref{gff2}.

Our goal is to find the BCFT data for 3d i.e. $\epsilon=1$. We identify $\phi$ as $\sigma$ i.e.  the lightest $Z_2$ odd bulk operator in the Ising spectrum. We take its scaling dimension from the bootstrap result \cite{simmons2017lightcone, Chang:2024whx}
\begin{equation}
\Delta_\phi=\Delta_\sigma=0.518149\,.
\end{equation}
The bulk data in the BCFT setup is fixed by the scalars appearing in  $\sigma\times \sigma$ OPE. They are known from numerical bootstrap as follows \cite{simmons2017lightcone, Chang:2024whx}
\begin{align}\label{Isingbulk-dimension}
    \Delta_0&=\Delta_{\varepsilon}=1.41262\,,&  &\Delta_1=\Delta_{\varepsilon'}=3.8296\,,\nonumber\\ \Delta_2&=6.895\,, & &\Delta_3=7.253\,.
\end{align}
We truncate at 4 operators as the next scalar operator has a much higher dimension ($>10$) \cite{Ghosh:2026dse} and with a much smaller OPE coefficient $\lambda_\Delta$. 

\begin{table*}[t]
\small
\centering
\caption{Boundary spectrum for $\epsilon = 1$ compared against known lattice/MC and bootstrap results where available. We have $\Delta_\phi=\Delta_\sigma = 0.518149$. The quantities $\hat\Delta_0$, $\mu_{\Delta_0}$ ($\equiv b_{\phi\sigma}$) and $(a\lambda)_{\Delta_0}$ are known from Lattice Monte-Carlo \cite{Przetakiewicz2025}. The italicized value indicates input.  We used $\lambda_{\sigma\sigma\epsilon}=1.0518537(41)$ (conformal bootstrap, \cite{Simmons_Duffin_2017}).}
\label{tab:eps1_boundary_spectrum_compared}
\begin{tabular}{|c|c|c|c|c|c|c|c|}
\hline
$n$ & $\hat\Delta_n$ & $\hat\Delta_n$ (lattice) & $\mu^2_{\Delta_n}$ & $\mu_{\Delta_n}$ &  $\mu_{\Delta_n}$ (lattice) & $(a\lambda)_{\Delta_n}$ &  $(a\lambda)_{\Delta_n}$ (lattice)\\
\hline
0 & $\textit{0.3531(3)}$    & $\textit{0.3531(3)}$  & $2.0625(14)$              & $1.43614(47)$   & $1.435(3)$ & $1.2212(18)$   & $1.2202(42)$ \\
1 & $2.5254(27)$      & ---          & $0.024888(95)$            & $0.15776(30)$   & ---        & $0.08106(38)$  & --- \\
2 & $4.441(16)$        & ---          & $3.4780(19)\times10^{-4}$  & $0.018649(50)$ & ---        & $0.005062(23)$ & --- \\
3 & $6.4334(50)$       & ---          & $1.23(11)\times10^{-5}$    & $0.00350(16)$   & ---        & $-0.002414(29)$& --- \\
4 & $8.51569(86)$       & ---          & $(5.41\pm0.12)\times10^{-7}$    & $0.000736(82)$  & ---        & ---             & --- \\
\hline
\end{tabular}
\end{table*}

\begin{figure*}[t]
  \centering
  \begin{subfigure}[b]{0.32\textwidth}
    \centering
    \includegraphics[width=\textwidth]{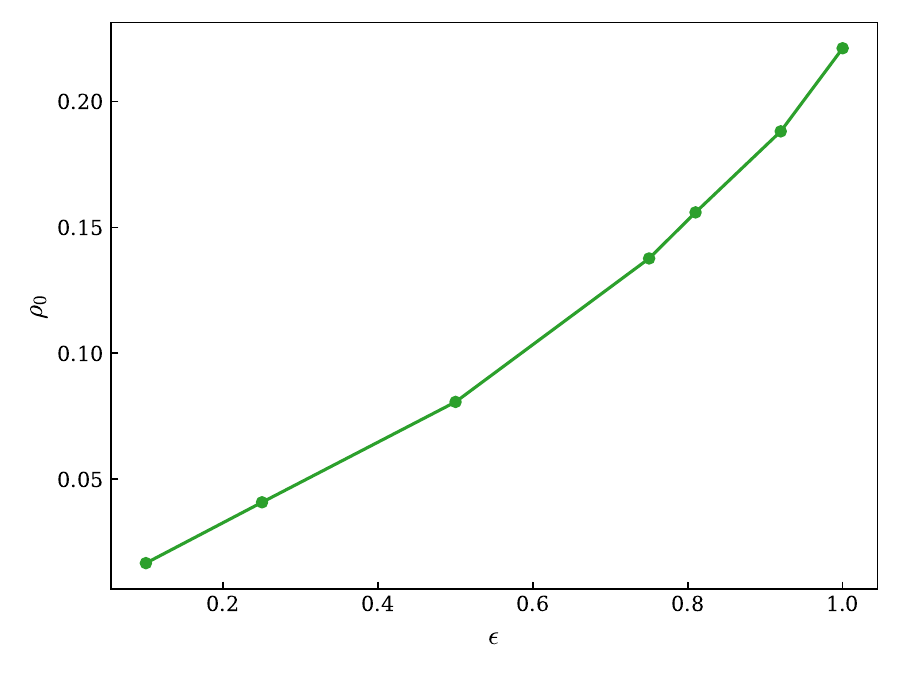}
    \label{fig:rho0vseps}
  \end{subfigure}
  \hfill
  \begin{subfigure}[b]{0.32\textwidth}
    \centering
    \includegraphics[width=\textwidth]{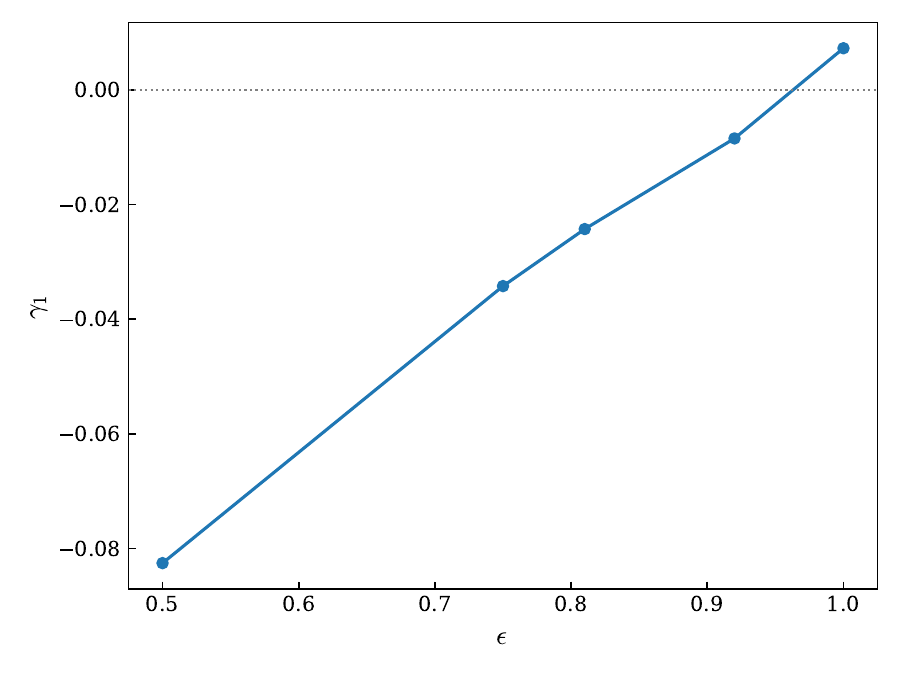}
    \label{fig:gamma1vseps}
  \end{subfigure}
  \hfill
  \begin{subfigure}[b]{0.32\textwidth}
    \centering
    \includegraphics[width=\textwidth]{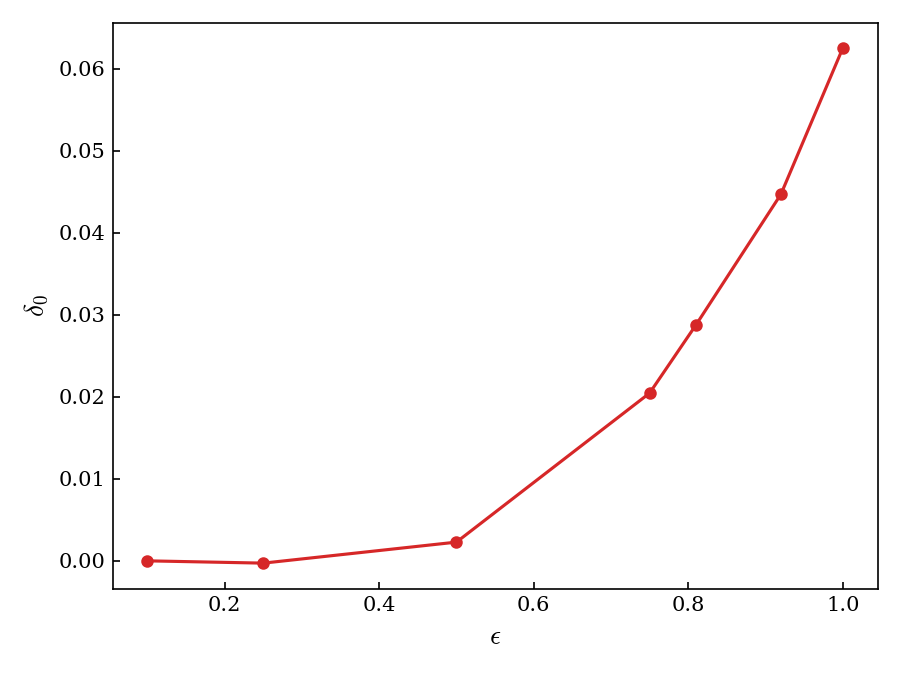}
    \label{fig:delta0vseps}
  \end{subfigure}
  \caption{The variations of $\rho_0$, $\gamma_1$ and $\delta_0$ vs $\epsilon$. The variations connect smoothly to $\epsilon=1$. The $\gamma_1$ is too small and unstable below $\epsilon=0.5$. }
  \label{fig:rhovseps}
\end{figure*}

We would like to use the truncated PB  to get the 3d results by treating 
the Wilson-Fisher BCFT as the starting point. So we will change $d$ i.e. increase $\epsilon$ from near 0 (free) to 1 step-wise as well as evolve the bulk data appropriately and run the optimization in each step. In order to get the best possible outcome it is useful to  incorporate the known 3d values in the bulk data evolution as the input  for each $0<\epsilon< 1$. 
For $\Delta_\phi$ this is achieved by a Pad\'e $[2,2]$ approximation:
\begin{equation}\label{Deltaphi-Ising}
    \Delta_\phi=1-\frac{3 \epsilon  (37 \epsilon +936)}{111 \epsilon ^2+326 \epsilon +5616}\,.
\end{equation}
This reduces to the perturbative expansion for $\epsilon\ll 1$ and is $\approx 0.518$ as $\epsilon\to 1$ \footnote{There is nothing sacred about this particular approximation. We are allowed to use a different one as long as it connects the two limits of $\epsilon$ smoothly}.
In order to connect the 3d bulk OPE spectrum to  4d  smoothly, we associate them to the free theory bulk operators as follows:
\begin{align}
\Delta_0\to\phi^2\,,\, \Delta_1\to\phi^4\,,\,\Delta_2\to\phi^6\,, \, \Delta_3\to(\partial\phi)^4\,.
\end{align}
The $\phi^2, \phi^4, \phi^6$  are simply the lightest $Z_2$ even operators, with free theory dimension $2, 4, 6$ respectively. The one-loop Wilson-Fisher dimension of $\phi^n$ is $n+ \epsilon n(n-4)/6$ \cite{rychkov2015epsilonexpansionconformalfieldtheory, Kehrein_1994}. 
At $\Delta=8$ there are two degenerate operators : $(\partial\phi)^4$ and $\phi^8$. We denote as $(\partial\phi)^4$ a composite primary with 4 $\phi$'s dressed with 4 derivatives - we choose it over $\phi^8$ because its one loop corrected dimension ($8-\frac{8}{9}\epsilon$) is lower \cite{Kehrein_1994}. In order to connect these four operators smoothly to 3d we use Pad\'e $[1,1]$ approximations. So for $0<\epsilon< 1$ we take the bulk dimensions to be
\begin{align}\label{Delta-bulk-Ising}
    &\Delta_0=2-\frac{\epsilon  (293 \epsilon +828)}{658 \epsilon +1242}\,, & \ & \Delta_1=4+\frac{9 \epsilon ^2}{9-59 \epsilon }\,,\nonumber\\
    &\Delta_2=6+\frac{112 \epsilon }{69 \epsilon +56}\,, & \ & \Delta_3=8-\frac{24 \epsilon}{5 \epsilon+27}\,.
\end{align}
Again these expressions are chosen such that they reduce to the perturbative Wilson-Fisher dimensions for small $\epsilon$ but are close to the values in \eqref{Isingbulk-dimension} at $\epsilon=1$.

Finally let us discuss the boundary operator $\hat{\phi}$. In our setup we will use its dimension $\hat{\Delta}_{\phi}:=\hat{\Delta}_0$ as an input too. In $d=3$ we will borrow its value $\hat\Delta_\phi=\hat{\Delta}_\sigma=0.3531(3)$ from the lattice Monte-Carlo analysis ~\cite{Przetakiewicz2025}. 
For general $0<\epsilon<1$  we employ the Pad\'e $[2,1]$ approximation to connect the above numerical value to the small $\epsilon$ result \eqref{bdyWF} - and write
\begin{equation}\label{bdyDphi}
    \hat\Delta_\phi=1+\frac{(4248-1013 \epsilon ) \epsilon }{1372 \epsilon -6372}\,.
\end{equation}

The rest of the BCFT data, for general $\epsilon$, are parameterized as follows:
\begin{align}\label{bulk-parameter-Ising}
&(a\lambda)_{\Delta_m} = (a\lambda)_{+,m}^{\text{gff}}+\rho_m(\epsilon)\,, \  &m=0,1,\cdots m_{\text{max}}\nonumber\\
&\hat{\Delta}_n=\Delta_\phi+2n+\gamma_n(\epsilon)\,, \  & n=0,1,\cdots n_{\text{max}}\,,\notag\,\\
&\mu_{\hat\Delta_n}^2=\mu^{\text{gff}}_{+,n}+\delta_n(\epsilon)\,. &
\end{align}
Obviously $m_{\text{max}}=3$ and we choose $n_{\text{max}}=4$. We have $\gamma_0$ fixed as an input.
The other parameters  $\rho_n(\epsilon), \gamma_m(\epsilon), \delta_n(\epsilon)$ will be computed numerically for various $\epsilon$ choices. Note that they must satisfy $\rho_m(\epsilon), \gamma_n(\epsilon), \delta_n(\epsilon)\to 0$ as  $\epsilon\to 0$. Furthermore, from \eqref{bulkWF} and \eqref{bdyWF} we also know that 
\begin{equation}
    \frac{\rho_0(\epsilon)} {\epsilon}\to \frac 16\,, 
    \ \frac{\delta_0(\epsilon)}{ \epsilon}\to 0\, \mbox{ \ \ as $\epsilon\to 0$}\,.
\end{equation}  

\paragraph{Results:} We evolve $\epsilon$ in steps with values from the set: $E=\{0.1,0.25,0.5,0.75,0.82, 0.92, 1\}$. For every $\epsilon$ we define the loss function like before:  
\begin{equation}
    \mathcal{F} = \sum_{n=0}^{n_{\text{max}}=4} a_n A_n^2 + \sum_{n=1}^{n_{\text{max}}=4} b_n B_n^2 + \sum_{m=0}^{m_{\text{max}}=3} t_m T_m^2\,.
    \label{eq:lossfunction-Ising}
\end{equation}
The number of $\alpha_n,\beta_n$ and $\theta_m$ sum rules coorespond to the number of variables $\delta_n,\gamma_n$ and $\rho_m$ to determine. So we fix $m_{\text{max}}=3$, $ n_{\text{max}}= 4$.

Each sum rule is evaluated using \eqref{Deltaphi-Ising} and \eqref{Delta-bulk-Ising} as bulk external $\Delta_\phi$ and bulk exchange $\Delta$  dimensions respectively, and \eqref{bdyDphi} as the leading boundary exchange $\hat\Delta_\phi$.  For $\epsilon=E_1=0.1$ we warm start with all parameters $\delta_n,\gamma_n, \rho_m=0$\,. For $\epsilon=E_{i}$ ($i>1$) we use the minimal spectrum of $\epsilon=E_{i-1}$ to warm start.

We show the leading OPE data: $\rho_0$, $\gamma_1$ and $\delta_0$ in Fig. \ref{fig:rhovseps}. The smooth variation with $\epsilon$ confirms we have tracked the same family of solutions. The results for $\epsilon=1$ are compiled in Table \ref{tab:eps1_boundary_spectrum_compared}. Error estimates are made with controlled random values in initializations, and strengths $a_n, b_n, t_m$  in \eqref{eq:lossfunction-Ising}. We find $\mu_{\hat\Delta_0}$ and $(a\lambda)_{\Delta_0}$ to be remarkably close to the Lattice values \cite{Przetakiewicz2025}. Other results in Table \ref{tab:eps1_boundary_spectrum_compared} are new. 

One last thing to point out is that the boundary spectrum we find is the analogue of an `extremal spectrum'. Even though there is no positivity we have tracked a single tower of GFF boundary states \cite{Ghosh:2025sic}. One cannot see other towers by bootstrapping a single BCFT correlator. However extremal spectra are often good approximations to the actual ones especially for the low lying OPE data \cite{Ghosh:2026dse}. For our results the same may be argued due to the excellent agreement of the leading BCFT data with lattice results.

\section{Conclusion}

We extended the truncated PB to boundary CFTs. First, we found a flow of GFF BCFTs between two different boundary conditions, triggered by a deformation on a co-dimension 1 AdS brane. Next, we bootstrapped the special surface transition of 3d Ising and found a set of new BCFT data. In both applications we obtained a family of solutions that smoothly interpolates between a free theory (GFF) with Neumann B. C. and a nonperturbative spectrum. 
Our analysis shows that any BCFT crossing solution can be identified by a deformation smoothly connecting it to the free theory.


It will be interesting to extend our setup to general defects. The analogue of analytic functionals should be dispersive sum rules for defects \cite{Barrat:2022psm, Bianchi_2023}. 
Another exciting direction is to generalize the framework for  three or higher point  functions \cite{Buric:2020zea}. Interactions of bulk fields with boundaries in the context of holography lead to locality conditions \cite{Levine:2023ywq, Levine:2024wqn} that are important to identify QFTs in AdS \cite{Antunes:2024hrt, Antunes:2021abs}.

\textbf{Acknowledgement:} We thank Ant\'onio Antunes, Koushik Ray, Aninda Sinha and Philine van Vliet for discussions and comments on the draft. AK is supported by ANRF grant ANRF/ARG/2025/001338/PS. 
\section{Appendix}

In this appendix, we review the explicit form of bulk and boundary functionals from ~\cite{Kaviraj2019}. We focus on only the ones corresponding to the GFF basis with Neumann boundary conditions, that we used in the main text. 
The generic form of all the functionals is given by:
\begin{equation}\label{functional form}
\omega[F(z)] :=  \int_1^\infty dz\, h_\omega(z)\,\mathrm{Im}[F(z)]\,.
\end{equation}
Here $\omega=\alpha_n, \beta_n$ or $\theta_n$. The forms of the kernels $h_\omega$ are determined by requiring a convergent action of the integral on a correlator $F=\mathcal{F}(z)$ (which depends on the boundedness property \eqref{Regge}) and also on its bulk and boundary block expansions.

For the boundary functionals it is useful to define 
\begin{align}
&h_{m,k}(z|\Delta_\phi) := -\frac{4^{k-m}\pi^{3/2}\,\Gamma(1+\Delta_\phi+m)}{\Gamma(1+m-k)}\,
z^{-\frac{3}{2}-k} \notag \\
& \times {}_2\tilde{F}_1\!\left(\mbox{$k-m-\frac{1}{2},\,\frac{3}{2}+m-k;\,\frac{1}{2}+\Delta_\phi+k;\,\frac{1}{z}$}\right)\,.
\end{align}
The kernels for the two lowest boundary functionals $\beta_1$ and $\alpha_0$ are given by
\begin{align*}
&h_{\beta_1}(z|\Delta_\phi)  = h_{1,0}(z|\Delta_\phi)  = z^{-1} h_{0,0}(z|\Delta_\phi + 1) \\
&h_{\alpha_0}(z|\Delta_\phi) = \frac{8}{\Delta_\phi(1+\Delta_\phi)} \left( h_{1,0}(z|\Delta_\phi) + h_{1,1}(z|\Delta_\phi) \right)  
\end{align*}
Higher boundary functionals are  obtained by defining `shifted functionals' $s\beta_n$ and $s\alpha_n$. They have the same form as \eqref{functional form} with the respective kernels given as follows: 
\begin{align}
h_{s\beta_n}(z) = z^{1-2n} h_{0,0}(z|\Delta_\phi - 1 + 2n) \\
h_{s\alpha_n}(z) = z^{1-2n} h_{\alpha}(z|\Delta_\phi - 1 + 2n) 
\end{align}
for $n\geq 1$. Here we have defined:
\begin{align*}
h_{\alpha}(z) := \frac{8\left( h_{2,2}(z) - h_{2,0}(z) + n_{\Delta_\phi} h_{0,0}(z) \right) }{(1+\Delta_\phi)(2+\Delta_\phi)}  
\end{align*}
\begin{align*}
n_{\Delta_\phi} := \frac{(\Delta_\phi + 1)(\Delta_\phi + 2)}{32} \left(1 + 8\log(2) - 4 H_{\Delta_\phi}\right)\,.
\end{align*}
The shifted functionals $s\beta_n$ and $s\alpha_n$ are linear combinations of a finite set of the desired functionals $\beta_m, \alpha_m$ with $m\leq n$. The first few such relations are: $s\beta_1=\beta_1$, $s\alpha_0=a_{00}\alpha_1+b_{01}\beta_1$, $s\alpha_1=a_{11}\alpha_1+b_{11}\beta_1+a_{10}\alpha_0$. Higher $n$'s are similar. The $\beta_n$ and $\alpha_n$  can be obtained from $s\beta_n$ and $s\alpha_n$ by requiring the orthogonality conditions \eqref{ortho1}.

Now let us present the bulk functionals $\theta_n$. For them we define the function
\begin{align*}
h_{\theta}(z|\Delta_\phi) := N_{\Delta_\phi} \sqrt{z} \Bigg[ \frac{(\Delta_\phi - \frac{1}{2})}{2(z-1)} \, {}_2\tilde{F}_1\left(\mbox{$-\frac{1}{2}, \frac{3}{2}, \frac{1}{2}+\Delta_\phi; \frac{1}{z}$}\right) \\
+\, {}_2\tilde{F}_1\left(\mbox{$-\frac{1}{2}, \frac{1}{2}, \Delta_\phi - \frac{3}{2}; \frac{1}{z}$}\right)
- {}_2\tilde{F}_1\left(\mbox{$-\frac{1}{2}, -\frac{1}{2}, \Delta_\phi - \frac{3}{2}; \frac{1}{z}$}\right) \Bigg].
\end{align*}
with
\begin{align*}
N_{\Delta_\phi} := \frac{4\Gamma(\Delta_\phi - 1)\Gamma(\Delta_\phi + 1)}{\pi \Gamma\left(\frac{1}{2} + \Delta_\phi\right)}.
\end{align*}
The bulk functionals are also defined in terms of shifted functionals $s\theta_n$. The respective kernels are given by
\begin{align*}
\quad h_{s\theta_n}(z) = \frac{z}{(z-1)^n} h_{\theta}(z|\Delta_\phi - 1 + 2n), \qquad n \geq 1 
\end{align*}
For $n=0$ the above kernel is incompatible  with convergence of the functional action.
So for $s\theta_0=\theta_0$ we define a kernel separately as follows:
\begin{align*}
\quad h_{\theta_0}(z) = z \left[ h_{\theta}(z|\Delta_\phi - 1) - \frac{\sqrt{\pi}\,\Gamma(\Delta_\phi)}{\Gamma\left(\Delta_\phi + \frac{1}{2}\right)} \, h_{0,0}(z|\Delta_\phi - 1) \right] \,.
\end{align*}
The shifted functional $s\theta_n$ is a linear combination of $\theta_m$ with $m\leq n$, e.g. $s\theta_1=t_{11}\theta_1+t_{10}\theta_0$. So $\theta_n$ is recovered from $s\theta_n$ using orthogonality \eqref{ortho1}.

\bibliographystyle{apsrev4-2}
\bibliography{references}

@article{Cardy1984,
title = {Conformal invariance and surface critical behavior},
journal = {Nuclear Physics B},
volume = {240},
number = {4},
pages = {514-532},
year = {1984},
issn = {0550-3213},
doi = {https://doi.org/10.1016/0550-3213(84)90241-4},
url = {https://www.sciencedirect.com/science/article/pii/0550321384902414},
author = {John L. Cardy}
}

@article{Liendo_2012,
   title={The bootstrap program for boundary CFT d},
   volume={2013},
   ISSN={1029-8479},
   url={http://dx.doi.org/10.1007/JHEP07(2013)113},
   DOI={10.1007/jhep07(2013)113},
   number={7},
   journal={Journal of High Energy Physics},
   publisher={Springer Science and Business Media LLC},
   author={Liendo, Pedro and Rastelli, Leonardo and van Rees, Balt C.},
   year={2013},
   month=July }

@article{Bissi2018,
   title={Analytic bootstrap for boundary CFT},
   volume={2019},
   ISSN={1029-8479},
   url={http://dx.doi.org/10.1007/JHEP01(2019)010},
   DOI={10.1007/jhep01(2019)010},
   number={1},
   journal={Journal of High Energy Physics},
   publisher={Springer Science and Business Media LLC},
   author={Bissi, Agnese and Hansen, Tobias and Söderberg, Alexander},
   year={2019},
   month=Jan }

@article{Kaviraj2019,
  author  = {Kaviraj, A. and Paulos, M. F.},
  title   = {The Functional Bootstrap for Boundary CFT},
  journal = {JHEP},
  volume  = {04},
  pages   = {135},
  year    = {2020},
  eprint  = {1812.04034},
  archivePrefix = {arXiv},
  primaryClass = {hep-th}
}

@article{Przetakiewicz2025,
  author  = {Przetakiewicz, Dorian and Wessel, Stefan and Parisen Toldin, Francesco},
  title   = {Boundary Operator Product Expansion Coefficients of the Three-dimensional Ising Universality Class},
  journal = {Phys. Rev. Research},
  volume  = {7},
  pages   = {L032051},
  year    = {2025},
  eprint  = {2502.14965},
  archivePrefix = {arXiv},
  primaryClass = {cond-mat.stat-mech}
}

@article{Simmons_Duffin_2017,
   title={The lightcone bootstrap and the spectrum of the 3d Ising CFT},
   volume={2017},
   ISSN={1029-8479},
   url={http://dx.doi.org/10.1007/JHEP03(2017)086},
   DOI={10.1007/jhep03(2017)086},
   number={3},
   journal={Journal of High Energy Physics},
   publisher={Springer Science and Business Media LLC},
   author={Simmons-Duffin, David},
   year={2017},
   month=Mar }

@article{Mazac:2018biw,
    author = "Maz{\'a}{\v{c}}, Dalimil and Rastelli, Leonardo and Zhou, Xinan",
    title = "{An analytic approach to BCFT$_{d}$}",
    eprint = "1812.09314",
    archivePrefix = "arXiv",
    primaryClass = "hep-th",
    reportNumber = "PUPT-2580; YITP-44",
    doi = "10.1007/JHEP12(2019)004",
    journal = "JHEP",
    volume = "12",
    pages = "004",
    year = "2019"
}

@article{Zan:2019fkw,
    author = "Paulos, Miguel F. and Zan, Bernardo",
    title = "{A functional approach to the numerical conformal bootstrap}",
    eprint = "1904.03193",
    archivePrefix = "arXiv",
    primaryClass = "hep-th",
    doi = "10.1007/JHEP09(2020)006",
    journal = "JHEP",
    volume = "09",
    pages = "006",
    year = "2020"
}

@article{Ghosh:2025sic,
    author = "Ghosh, Kausik and Paulos, Miguel F. and Suchel, No{\'e}",
    title = "{Solving 1D crossing and QFT$_{2}$/CFT$_{1}$}",
    eprint = "2503.22798",
    archivePrefix = "arXiv",
    primaryClass = "hep-th",
    doi = "10.1007/JHEP04(2026)119",
    journal = "JHEP",
    volume = "04",
    pages = "119",
    year = "2026"
}

@article{Padayasi:2021sik,
    author = "Padayasi, Jaychandran and Krishnan, Abijith and Metlitski, Max A. and Gruzberg, Ilya A. and Meineri, Marco",
    title = "{The extraordinary boundary transition in the 3d O(N) model via conformal bootstrap}",
    eprint = "2111.03071",
    archivePrefix = "arXiv",
    primaryClass = "cond-mat.stat-mech",
    doi = "10.21468/SciPostPhys.12.6.190",
    journal = "SciPost Phys.",
    volume = "12",
    number = "6",
    pages = "190",
    year = "2022"
}

@article{simmons2017lightcone,
  title={The lightcone bootstrap and the spectrum of the 3d Ising CFT},
  author={Simmons-Duffin, David},
  journal={Journal of High Energy Physics},
  volume={2017},
  number={3},
  pages={1--85},
  year={2017},
  publisher={Springer}
}

@article{Antunes:2021abs,
    author = "Antunes, Ant{\'o}nio and Costa, Miguel S. and Penedones, Jo{\~a}o and Salgarkar, Aaditya and van Rees, Balt C.",
    title = "{Towards bootstrapping RG flows: sine-Gordon in AdS}",
    eprint = "2109.13261",
    archivePrefix = "arXiv",
    primaryClass = "hep-th",
    doi = "10.1007/JHEP12(2021)094",
    journal = "JHEP",
    volume = "12",
    pages = "094",
    year = "2021"
}

@article{Antunes:2024hrt,
    author = "Antunes, Ant{\'o}nio and Lauria, Edoardo and van Rees, Balt C.",
    title = "{A bootstrap study of minimal model deformations}",
    eprint = "2401.06818",
    archivePrefix = "arXiv",
    primaryClass = "hep-th",
    reportNumber = "DESY-23-141",
    doi = "10.1007/JHEP05(2024)027",
    journal = "JHEP",
    volume = "05",
    pages = "027",
    year = "2024"
}

@article{Levine:2023ywq,
    author = "Levine, Nat and Paulos, Miguel F.",
    title = "{Bootstrapping bulk locality. Part I: Sum rules for AdS form factors}",
    eprint = "2305.07078",
    archivePrefix = "arXiv",
    primaryClass = "hep-th",
    doi = "10.1007/JHEP01(2024)049",
    journal = "JHEP",
    volume = "01",
    pages = "049",
    year = "2024"
}

@article{Levine:2024wqn,
    author = "Levine, Nat and Paulos, Miguel F.",
    title = "{Bootstrapping bulk locality. Part II: Interacting functionals}",
    eprint = "2408.00572",
    journal="",
    archivePrefix = "arXiv",
    primaryClass = "hep-th",
    month = "8",
    year = "2024"
}

@article{Ghosh:2026dse,
    author = "Ghosh, Kausik and Zheng, Zechuan",
    title = "{Bootstrapping 3D Conformal Field Theories with Product Analytic Functionals}",
    eprint = "2608.06497",
    archivePrefix = "arXiv",
    primaryClass = "hep-th",
    journal="",
    month = "8",
    year = "2026"
}

@article{Bissi_2019,
   title={Analytic bootstrap for boundary CFT},
   volume={2019},
   ISSN={1029-8479},
   url={http://dx.doi.org/10.1007/JHEP01(2019)010},
   DOI={10.1007/jhep01(2019)010},
   number={1},
   journal={Journal of High Energy Physics},
   publisher={Springer Science and Business Media LLC},
   author={Bissi, Agnese and Hansen, Tobias and Söderberg, Alexander},
   year={2019},
   month=Jan }

@article{Kehrein_1994,
   title={The structure of the spectrum of anomalous dimensions in the N-vector model in 4 − ϵ dimensions},
   volume={424},
   ISSN={0550-3213},
   url={http://dx.doi.org/10.1016/0550-3213(94)90406-5},
   DOI={10.1016/0550-3213(94)90406-5},
   number={3},
   journal={Nuclear Physics B},
   publisher={Elsevier BV},
   author={Kehrein, Stefan K. and Wegner, Franz},
   year={1994},
   month=Aug, pages={521–546} }

@misc{rychkov2015epsilonexpansionconformalfieldtheory,
      title={The Epsilon-Expansion from Conformal Field Theory}, 
      author={Slava Rychkov and Zhong Ming Tan},
      year={2015},
      eprint={1505.00963},
      archivePrefix={arXiv},
      primaryClass={hep-th},
      url={https://arxiv.org/abs/1505.00963}, 
}

@article{Chang:2024whx,
    author = "Chang, Cyuan-Han and Dommes, Vasiliy and Erramilli, Rajeev S. and Homrich, Alexandre and Kravchuk, Petr and Liu, Aike and Mitchell, Matthew S. and Poland, David and Simmons-Duffin, David",
    title = "{Bootstrapping the 3d Ising stress tensor}",
    eprint = "2411.15300",
    archivePrefix = "arXiv",
    primaryClass = "hep-th",
    reportNumber = "CALT-TH 2024-047",
    doi = "10.1007/JHEP03(2025)136",
    journal = "JHEP",
    volume = "03",
    pages = "136",
    year = "2025"
}

@book{Cardy_1996, place={Cambridge}, series={Cambridge Lecture Notes in Physics}, title={Scaling and Renormalization in Statistical Physics}, publisher={Cambridge University Press}, author={Cardy, John}, year={1996}, collection={Cambridge Lecture Notes in Physics}}

@article{Dedushenko:2024nwi,
    author = "Dedushenko, Mykola",
    title = "{Ising BCFT from Fuzzy Hemisphere}",
    eprint = "2407.15948",
    archivePrefix = "arXiv",
    primaryClass = "hep-th",
    month = "7",
     journal = "",
    year = "2024"
}

@article{Barrat:2022psm,
    author = "Barrat, Julien and Gimenez-Grau, Aleix and Liendo, Pedro",
    title = "{A dispersion relation for defect CFT}",
    eprint = "2205.09765",
    archivePrefix = "arXiv",
    primaryClass = "hep-th",
    reportNumber = "HU-EP-22/18-RTG",
    doi = "10.1007/JHEP02(2023)255",
    journal = "JHEP",
    volume = "02",
    pages = "255",
    year = "2023"
}

@article{Bianchi_2023,
   title="Conformal dispersion relations for defects and boundaries",
   volume="15",
   ISSN="2542-4653",
   url="http://dx.doi.org/10.21468/SciPostPhys.15.2.055",
   DOI="10.21468/scipostphys.15.2.055",
   number="2",
   journal="SciPost Physics",
   publisher="Stichting SciPost",
   author="Bianchi, Lorenzo and Bonomi, Davide",
   year="2023",
   month=Aug }

@article{Buric:2020zea,
    author = "Buri{\'c}, Ilija and Schomerus, Volker",
    title = "{Defect Conformal Blocks from Appell Functions}",
    eprint = "2012.12489",
    archivePrefix = "arXiv",
    primaryClass = "hep-th",
    reportNumber = "DESY-20-206, DESY 20-206",
    doi = "10.1007/JHEP05(2021)007",
    journal = "JHEP",
    volume = "05",
    pages = "007",
    year = "2021"
}

@article{Feng:2026iii,
    author = "Feng, Jiechao and Wang, Taige",
    title = "{Studying 3D O(N) Surface CFT on the Fuzzy Sphere}",
    eprint = "2604.21091",
    archivePrefix = "arXiv",
    primaryClass = "cond-mat.str-el",
    journal= " ",
    month = "4",
    year = "2026"
}

@article{Diatlyk:2026eta,
    author = "Diatlyk, Oleksandr and Giombi, Simone and Sun, Zimo",
    title = "{Boundary criticality in the Gross-Neveu-Yukawa model at higher orders}",
    eprint = "2606.07510",
    archivePrefix = "arXiv",
    primaryClass = "hep-th",
    journal= " ",
    month = "6",
    year = "2026"
}

@article{Hu:2025yrs,
    author = "Hu, Runzhe and Li, Wenliang",
    title = "{Accurate Boundary Bootstrap for the Three-Dimensional O(N) Normal Universality Class}",
    eprint = "2508.20854",
    archivePrefix = "arXiv",
    primaryClass = "hep-th",
    doi = "10.1103/hnd3-636j",
    journal = "Phys. Rev. Lett.",
    volume = "136",
    number = "22",
    pages = "221601",
    year = "2026"
}

@article{Guo:2026vmq,
    author = "Guo, Yongwei and Li, Wenliang",
    title = "{}",
    eprint = "2605.16119",
    archivePrefix = "arXiv",
    primaryClass = "hep-th",
    journal= " ",
    month = "5",
    year = "2026"
}

@article{Kangshabanik:2026onk,
    author = "Kangshabanik, Kaushik and Kaviraj, Apratim and Lal, Shailesh",
    title = "{Truncated Polyakov bootstrap}",
    eprint = "2609.01732",
    archivePrefix = "arXiv",
    journal = "",
    primaryClass = "hep-th",
    month = "9",
    year = "2026"
}

@article{Shpot:2019iwk,
    author = "Shpot, M. A.",
    title = "{Boundary conformal field theory at the extraordinary transition: The layer susceptibility to $O(\varepsilon)$}",
    eprint = "1912.03021",
    archivePrefix = "arXiv",
    primaryClass = "hep-th",
    doi = "10.1007/JHEP01(2021)055",
    journal = "JHEP",
    volume = "01",
    pages = "055",
    year = "2021"
}

@article{Giombi:2019enr,
    author = "Giombi, Simone and Khanchandani, Himanshu",
    title = "{$O(N)$ models with boundary interactions and their long range generalizations}",
    eprint = "1912.08169",
    archivePrefix = "arXiv",
    primaryClass = "hep-th",
    reportNumber = "PUPT-2606",
    doi = "10.1007/JHEP08(2020)010",
    journal = "JHEP",
    volume = "08",
    number = "08",
    pages = "010",
    year = "2020"
}

@article{Dey:2020lwp,
    author = "Dey, Parijat and Hansen, Tobias and Shpot, Mykola",
    title = "{Operator expansions, layer susceptibility and two-point functions in BCFT}",
    eprint = "2006.11253",
    archivePrefix = "arXiv",
    primaryClass = "hep-th",
    reportNumber = "UUITP-21/20",
    doi = "10.1007/JHEP12(2020)051",
    journal = "JHEP",
    volume = "12",
    pages = "051",
    year = "2020"
}

@article{Gimenez-Grau:2020jvf,
    author = "Gimenez-Grau, Aleix and Liendo, Pedro and van Vliet, Philine",
    title = "{Superconformal boundaries in $4-\epsilon$ dimensions}",
    eprint = "2012.00018",
    archivePrefix = "arXiv",
    primaryClass = "hep-th",
    reportNumber = "DESY 20-215, DESY-20-215",
    doi = "10.1007/JHEP04(2021)167",
    journal = "JHEP",
    volume = "04",
    pages = "167",
    year = "2021"
}

@article{Diehl1981FieldtheoreticalAT,
  title={Field-theoretical approach to static critical phenomena in semi-infinite systems},
  author={Hans Werner Diehl and S Dietrich},
  journal={Zeitschrift f{\"u}r Physik B Condensed Matter},
  year={1981},
  volume={43},
  pages={281},
  url={https://api.semanticscholar.org/CorpusID:121245636}
}

@article{PhysRevLett.45.1581,
  title = {Critical Behavior of the $n$-Vector Model with a Free Surface},
  author = {Reeve, J. S. and Guttmann, A. J.},
  journal = {Phys. Rev. Lett.},
  volume = {45},
  issue = {19},
  pages = {1581--1583},
  numpages = {0},
  year = {1980},
  month = {Nov},
  publisher = {American Physical Society},
  doi = {10.1103/PhysRevLett.45.1581},
  url = {https://link.aps.org/doi/10.1103/PhysRevLett.45.1581}
}

@article{REEVE1981237,
title = {Renormalisation group calculation of the critical exponents of the special transition in semi-infinite systems},
journal = {Physics Letters A},
volume = {81},
number = {4},
pages = {237-238},
year = {1981},
issn = {0375-9601},
doi = {https://doi.org/10.1016/0375-9601(81)90250-4},
url = {https://www.sciencedirect.com/science/article/pii/0375960181902504},
author = {Jeffrey S. Reeve}
}

@article{PhysRevB.11.4533,
  title = {Critical phenomena in semi-infinite systems. I. $\ensuremath{\epsilon}$ expansion for positive extrapolation length},
  author = {Lubensky, T. C. and Rubin, Morton H.},
  journal = {Phys. Rev. B},
  volume = {11},
  issue = {11},
  pages = {4533--4546},
  numpages = {0},
  year = {1975},
  month = {Jun},
  publisher = {American Physical Society},
  doi = {10.1103/PhysRevB.11.4533},
  url = {https://link.aps.org/doi/10.1103/PhysRevB.11.4533}
}

@article{Behan_2020,
   title={Bootstrapping boundary-localized interactions},
   volume={2020},
   ISSN={1029-8479},
   url={http://dx.doi.org/10.1007/JHEP12(2020)182},
   DOI={10.1007/jhep12(2020)182},
   number={12},
   journal={Journal of High Energy Physics},
   publisher={Springer Science and Business Media LLC},
   author={Behan, Connor and Di Pietro, Lorenzo and Lauria, Edoardo and van Rees, Balt C.},
   year={2020},
   month=Dec }

@article{Behan:2021tcn,
    author = "Behan, Connor and Di Pietro, Lorenzo and Lauria, Edoardo and van Rees, Balt C.",
    title = "{Bootstrapping boundary-localized interactions II. Minimal models at the boundary}",
    eprint = "2111.04747",
    archivePrefix = "arXiv",
    primaryClass = "hep-th",
    doi = "10.1007/JHEP03(2022)146",
    journal = "JHEP",
    volume = "03",
    pages = "146",
    year = "2022"
}

@article{Banerjee:2026gce,
    author = "Banerjee, Pinaki and Dey, Parijat and Dhar, Dipyendu",
    title = "{Boundary conformal field theory, holography and bulk locality}",
    eprint = "2602.04223",
    archivePrefix = "arXiv",
    primaryClass = "hep-th",
    doi = "10.1007/JHEP05(2026)096",
    journal = "JHEP",
    volume = "05",
    pages = "096",
    year = "2026"
}

@article{Zhou_2025,
   title={Studying the 3d Ising surface CFTs on the fuzzy sphere},
   volume={18},
   ISSN={2542-4653},
   url={http://dx.doi.org/10.21468/SciPostPhys.18.1.031},
   DOI={10.21468/scipostphys.18.1.031},
   number={1},
   journal={SciPost Physics},
   publisher={Stichting SciPost},
   author={Zhou, Zheng and Zou, Yijian},
   year={2025},
   month=Jan }

@article{McAvity_1995,
   title={Conformal field theories near a boundary in general dimensions},
   volume={455},
   ISSN={0550-3213},
   url={http://dx.doi.org/10.1016/0550-3213(95)00476-9},
   DOI={10.1016/0550-3213(95)00476-9},
   number={3},
   journal={Nuclear Physics B},
   publisher={Elsevier BV},
   author={McAvity, D.M. and Osborn, H.},
   year={1995},
   month=Sept, pages={522–576} 
   }

@misc{sun2026analyticbootstraponboundary,
      title={Analytic Bootstrap for $O(N)$ Boundary Conformal Field Theories with Interacting Boundaries}, 
      author={Xinyu Sun and Shao-Kai Jian and Hong Yao},
      year={2026},
      eprint={2605.28933},
      archivePrefix={arXiv},
      primaryClass={hep-th},
      url={https://arxiv.org/abs/2605.28933}, 
}

@misc{guo2026boundaryanomalousdimensionsbcft,
      title={Boundary anomalous dimensions from BCFT: $\phi^{3}$ theories with a boundary and higher-derivative generalizations}, 
      author={Yongwei Guo and Wenliang Li},
      year={2026},
      eprint={2605.16119},
      archivePrefix={arXiv},
      primaryClass={hep-th},
      url={https://arxiv.org/abs/2605.16119}, 
}

@article{Guo:2025edk,
    author = "Guo, Yongwei and Li, Wenliang",
    title = "{Boundary anomalous dimensions from BCFT: O(N)-symmetric {\ensuremath{\phi}}2n theories with a boundary and higher-derivative generalizations}",
    eprint = "2504.16844",
    archivePrefix = "arXiv",
    primaryClass = "hep-th",
    doi = "10.1103/pqs4-hs43",
    journal = "Phys. Rev. D",
    volume = "112",
    number = "6",
    pages = "065001",
    year = "2025"
}

\end{document}